\documentclass[prl,aps,twocolumn,showpacs,superscriptaddress]{revtex4}
\usepackage{graphicx}
\usepackage{psfrag}
\usepackage{ae}
\usepackage{amsmath,amssymb}
\usepackage{fancyhdr}

\begin{document}
\date{\today}

\title{Ion specificity and the theory of stability of colloidal suspensions}

\author{Alexandre P. dos Santos}
\author{Yan Levin}
\affiliation{Instituto de F\'{\i}sica, Universidade Federal do Rio Grande do Sul, Caixa Postal 15051, CEP 91501-970,
Porto Alegre, RS, Brazil}

\begin{abstract}

A theory is presented which allow us to accurately calculate the critical coagulation concentration (CCC) of hydrophobic colloidal suspensions.
For positively charged particles the CCC's follow the Hofmeister (lyotropic) series.  For negatively charged
particles the series is reversed.  We find that strongly polarizable chaotropic anions are driven towards the colloidal surface by electrostatic and 
hydrophobic forces. Within approximately one ionic radius from the surface, the chaotropic anions loose part of their hydration sheath
and become strongly adsorbed.  The kosmotropic anions, on the other hand, are repelled from the hydrophobic surface.  
The theory is quantitatively
accurate without any adjustable parameters.   
We speculate that the same mechanism is responsible for the Hofmeister series that
governs stability of protein solutions.

\end{abstract}


\maketitle

All of biology is specific.  The ion channels which control the  electrolyte concentration
inside living cells  are specific to the  ions which they allow to pass. The
tertiary structure of proteins is sensitive to both the pH and to specific ions inside the solution.
It has been known for over a hundred years that while
some ions stabilize protein solutions, often denaturing them in the process, others
lead to protein precipitation.  In the fields as diverse as biophysics, biochemistry,
electrochemistry, and colloidal science,  ionic specificity 
has been known --- and puzzled over --- for
a very long time.   It has become 
known as the ``Hofmeister effect" or the ``lyotropic" series of electrolytes, depending on the field of
science.    
The traditional physical theories of electrolytes completely 
fail to account for the ion specificity.
The Debye-H\" uckel (DH) theory of electrolytes and the Onsager-Samaras theory of
surface tensions treat ions as hard spheres with a point charge located at the center~\cite{Le02}.
The cornerstone of colloidal science, the Derjaguin,
Landau, Verwey and Overbeek (DLVO) theory of stability of lyophobic colloidal suspensions
is based on an even simpler picture
of hard sphere-like colloidal particles interacting with the point-like ions
through Coulomb potential.  
The DLVO theory showed that 
the primary minimum of colloid-colloid interaction potential --- arising
from the mutual van der Waals (dispersion) attraction --- is not accessible at low
electrolyte  concentrations because of a large energy barrier.  
When the electrolyte concentration is raised above the critical
coagulation concentration (CCC), the barrier height drops down to zero leading to colloidal
flocculation and precipitation.  
The DLVO theory, however, predicts that the CCC 
should be the same for all monovalent electrolytes, which is clearly not the case~\cite{BoWi01_prl,LoJo03,LoSa08,PeOr10}.  In fact
it has been known for a long time that the effectiveness of electrolyte at precipitating hydrophobic
colloids follows the lyotropic series.  For positively charged particles the 
CCC concentration of sodium thiocyanide
is an order of magnitude lower than the CCC of sodium fluoride.  Even more
dramatic is the variation of the CCC with the sign of colloidal charge.  The CCC's predicted
by the DLVO theory are completely invariant under the colloidal charge reversal.  Therefore, changing
the charge from say $+0.04$ to $-0.04~\rm{C/m^2}$, results in exactly the same CCC.  This
is completely contradicted by the experiments which find that under the same
charge reversal, the CCC's can change by as much as two orders of magnitude for {\it exactly the
same electrolyte}.  Similar dramatic effects are observed for protein solutions for which
some electrolytes are found to ``salt-in"  while other ``salt-out" the {\it same} protein.
In view of the fundamental importance of ion specificity across so many different disciplines,
there has been a great effort to understand its physical 
mechanisms~\cite{CoWa85,Ga04,KuLo04,Co04,ZhCr06,PeRe07,ZaHa07,LuVr08,ScHo10}.
A successful theory should be able to quantitative predict the CCC's of colloidal suspensions and
shed new light on the specific ion effects so fundamental to modern biology.
Construction of such a theory is the subject of the present Letter.

Over the last twenty years there has been a growing realization that ionic polarizability  --- the
effective rigidity of the electronic charge distribution --- plays an important role in coding the 
ionic specificity~\cite{PeBe91,DaSm93,StBr99,Le09}.
The work on surface tension of electrolyte solutions showed 
that near an air-water interface ions can be divided into two classes: kosmotropes and chaotropes~\cite{LeDo09,DoDi10}.
The kosmotropes remain strongly hydrated, and are repelled from the interface.  On the other hand, chaotropes 
loose their hydration sheath and redistribute their electronic charge so that it remains mostly hydrated.  
This way chaotropes gain hydrophobic free energy at a small price in electrostatic self-energy.
For hard non-polarizable ions of the Debye-H\" uckel-Onsager-Samaras theory, the hydrophobic 
forces are just too weak to overcome the electrostatic
self-energy penalty of exposing the ionic charge to the low dielectric air environment, forcing
these ions to always stay in the bulk, contrary to simulations and experiments. 

\begin{figure}[h]
\begin{center}
\includegraphics[scale=0.2]{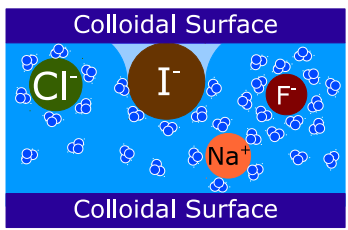}
\end{center}
\caption{Illustrative representation: while the chaotropic ions loose part of their hydration sheath near a hydrophobic 
surface and become adsorbed,  the kosmotropic ions are repelled from the surface by the hydration layer.}
\label{fig1}
\end{figure}

It is natural to ask if the same mechanism is also responsible for the ionic specificity of the CCCs
and for the stability of protein solutions. The similarity between the air-water
interface and a hydrophobe-water interface, however, is not so straight forward~\cite{Ch05}.
The simulations find that contrary to the air-water interface,  the  mean water density is actually {\it larger} near a
hydrophobic surface than in the bulk~\cite{GoJa09}.  On the other hand, the interfacial water 
is ``softer" than bulk water, so that the hydration layer is very compressible~\cite{GoJa09}.  In particular, the simulations find that 
the energy cost of placing
an ideal cavity of radius $a$ near a hydrophobic surface is half that of having it in the bulk~\cite{GoJa09}.  For small cavities, the bulk energy cost
scales with the volume~\cite{LuCh99,Ch05} as $U_{bulk}=\nu a^3$, where $\nu=0.3k_B T$~\cite{RaTr05}.  
Moving an ion to the  surface, therefore, gains $U_{cav}=-\frac{\nu}{2} a^3$ of hydrophobic energy.  
For kosmotropic ions this energy is too small to compensate the loss of the
hydration sheath and for exposing part of the ionic charge to the low dielectric environment of the surface.  For highly polarizable 
chaotropic ions the cavitational energy gain compensates the electrostatic energy cost, since these ions can shift most of their
electronic charge to still remain hydrated. Furthermore, if there are no water-molecules between an ion and a surface,
dispersion interactions --- resulting from quantum electromagnetic field fluctuations --- come into play.  
Dispersion forces are strongly attenuated by water.  For example, the Hamaker
constant for polystyrene in water is seven times lower than it is in vacuum~\cite{Ru89}. The London-Lifshitz theory 
predicts that the dispersion
interaction between an ion and a surface decays with the 
third power of separation and is proportional to the ionic polarizability. Although there is no exponential screening of 
dispersion force by electrolyte, in practice, it is very short ranged and is only relevant for 
chaotropic ions in almost a physical contact with a hydrophobic surface, see Fig.~\ref{fig1}.

We shall work in the grand-canonical ensemble.  The colloidal particles will be treated
as planar surfaces of surface charge density $\sigma$ with the electrolyte in between. To account for the curvature we
will use the Derjaguin approximation~\cite{Ru89}. The static 
dielectric constant of the medium (water) is $\epsilon_w$.  The Bjerrum length is defined as $\lambda_B=\beta q^2/\epsilon_w$, where $\beta=1/k_BT$,
and is $7.2$ \AA, in water at room temperature.
The system --- two plates with electrolyte in between --- is in contact with a salt reservoir at concentration $\rho_S$. 
The number of cations and anions $N_\pm$ per unit area in between the surfaces, as well as their spatial distribution will 
be calculated by minimizing the grand-potential function.  
Note, that we do not require the charge neutrality inside the system --- 
the external electrolyte will screen the excess charge.  
Of course, if the system is in vacuum,  the charge neutrality will be enforced by the formalism developed below.


The grand-potential function is
\begin{eqnarray}
\label{eq1}
\frac{\beta \Omega(L)}{A}= 
\int_{-L/2}^{L/2}dx \ \rho_+(x) \left[\ln{(\dfrac{\rho_+(x)}{\rho_S})}-1\right] + \nonumber \\
\int_{-L/2}^{L/2}dx \ \rho_-(x) \left[\ln{(\dfrac{\rho_-(x)}{\rho_S})}-1\right] + \nonumber \\
\beta\dfrac{\epsilon_w}{8\pi} \int_{-L/2}^{L/2}dx \ E(x)^2 + \nonumber\\
\beta \int_{-L/2}^{L/2}dx \ \rho_+(x)  U_+(x) +  \nonumber\\
\beta \int_{-L/2}^{L/2}dx \ \rho_-(x)  U_-(x) +\dfrac{\pi\beta}{2 \epsilon_w\kappa_w}\Delta^2\ + 2\rho_S L, \ \ \
\end{eqnarray}
where $A$ is the surface area, $x$ is the distance measured from midplane, $L$ is the separation between the plates, $\rho_{\pm}(x)$ 
are the ionic concentrations, $\Delta = qN_- -qN_+ - 2 \sigma$  is the deviation from the charge neutrality
inside the system, and  $U_{\pm}(x)$ are
the interaction potentials between the ions and the surfaces.  
The first two terms of eq.~(\ref{eq1}) are the entropic free energy,  the third term
is the electrostatic energy inside the system, the sixth term is the electrostatic penalty for violating the charge neutrality, 
and the last term accounts for the work that must be done against the pressure of reservoir.
The electrostatic penalty (sixth term) is calculated using the DH equation (linearized Poisson-Boltzmann equation) 
and accounts for the screening of the excess charge 
by the external reservoir. The inverse (screening) Debye length of the reservoir is $\kappa_w=\sqrt{8\pi \lambda_B \rho_S}$.  
The equilibrium values of $N_\pm$ and the density profiles $\rho_\pm(x)$
are determined by numerically minimizing the grand-potential function.
Using the Gauss law, the electric field is $E(x)=\dfrac{4\pi q}{\epsilon_w} \int_{0}^x \left[ \rho_+(x') -  \rho_-(x') \right] dx' $.
Minimizing the grand-potential function for a fixed $N_\pm$, we find
\begin{equation}
\label{eq3}
\rho_{\pm}(x)=\dfrac{N_{\pm}\Theta(\frac{L}{2}-a_{\pm}-x) e^{{\mp}\beta q \phi(x)+\beta U_{\pm}(x)}}{2 \int_{0}^{\frac{L}{2}-a_\pm} dx' e^{{\mp}\beta q \phi(x')+\beta U_{\pm}(x')}} \ ,
\end{equation}
where the electrostatic potential is  $\phi(x)=-\int_{0}^x dx'E(x')$,
$\Theta$ is the Heaviside step function, $a_{\pm}$ are the hydrated or unhydrated (bare) ionic radius, depending if the ion is a kosmotrope or a chaotrope. Substituting the ionic densities into Gauss' law, yields 
an integral equation for the electric field which is 
then solved numerically.  We proceed as follows: for a trial number of cations and anions,  $N_\pm$, 
we calculate the electric field and the ionic density profile. 
Once these are obtained, we substitute them back into the free energy functional (\ref{eq1}).  The trial values of $N_\pm$ are varied
until the minimum of the grand-potential is found.  In practice, we couple the integral equation solver to
a minimization subroutine.

The van der Waals interaction between two surfaces, is given by~\cite{Ru89}: $H(L)=-A_{pp}^w/12\pi L^2$, where $A_{pp}^w$ is the Hamaker constant for polystyrene in water~\cite{Ru89}. To account for the finite radius of colloidal particles, we use the Derjaguin approximation~\cite{Ru89}.  The total interaction
potential between the two latex spheres of radius $R_c$ is then $U_{tot}(L)=\pi R_c\int_L^{\infty}dl\left[\Omega(l)-\Omega(\infty)+H(l)\right] $.

\begin{figure}[h]
\begin{center}
\includegraphics[scale=0.2]{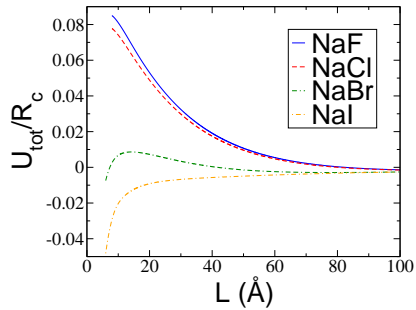}
\end{center}
\caption{Particle-particle interaction potentials in units of $k_BT/$\AA, for suspensions with sodium-halide salts. 
The colloidal surface charge is $0.04~\rm{C/m^2}$.  The electrolyte concentration is $\rho_S=20~\rm{mM}$. While suspensions containing
$NaF$, $NaCl$, and $NaBr$ at this concentration are stable -- there is an energy barrier for the primary minimum  -- 
suspension with $NaI$ is unstable.}
\label{fig2}
\end{figure}

We begin by studying the colloidal suspensions with sodium salts of kosmotropic anions, $IO_3^{-}$, $F^{-}$, $BrO_3^{-}$, $Cl^{-}$. 
For these salts both cations and  anions are strongly hydrated and are repelled
from the colloidal surface.  Since the static dielectric constant of latex is less than of water, in addition to 
the hard-core repulsion at contact, the ions also feel a repulsive charge-image interaction. 
To explicitly calculate this potential one needs to resum an infinite number of images.  This can be done using the 
theory developed in Ref.~\cite{LeMe01}, yielding the ion-image interaction potential $W_{im}(x)$.

For kosmotropic ions the latex-ion interaction potential has a particularly simple form  --- hard core repulsion at one hydrated ionic radius from each surface plus the ion-image interaction.  The hydrated radii were taken from Nightingale~\cite{Ni59} and are the same as used in our previous work on surface tensions~\cite{LeDo09,DoDi10}. The interaction potential between the two colloidal particles, $U_{tot}$, can now be calculated, Fig.~\ref{fig2}.  The 
potential has a primary minimum at $L=0$, followed by an energy barrier.  The CCC is defined as the concentration $\rho_S$ for which the barrier goes down to zero.
In Table~\ref{tab1} we present the calculated values of CCC for positively charged colloidal suspensions 
and compare the theoretical results with the experimental measurements. A very good agreement between the theory and
experiments is found {\it without any adjustable parameters}. To further test the accuracy of the theory we have calculated the CCC
for negatively charged colloidal particles with $\sigma=-0.061~\rm{C/m^2}$ in a suspension with $NaCl$~\cite{LoJo03}. In this case 
the experimental value of CCC was found to be $140~\rm{mM}$, while we obtain $134.2~\rm{mM}$.

For chaotropic anions the situation is significantly more complicated. As mentioned earlier these ions can shed their hydration sheath and adsorb to the colloidal
surface, gaining hydrophobic free energy.  In addition, diminished hydration leads to strong dispersion interaction between latex and
a chaotropic anion.  The London-Lifshitz theory predicts that the ion-particle dispersion potential has the form $U_{dis}(x)\approx B [\frac{1}{(L/2 + x)^3}+\frac{1}{(L/2 - x)^3}]$, where $B$ is the constant related to ionic
excess polarizability and the ionization potential, as well as the dielectric properties of polystyrene and water. 
Unfortunately these quantities are not known.  However,
we can get a reasonable approximation for $B$ using a simplified Hamaker theory~\cite{Ru89}. Within this approximation we find $B=2A_{\rm eff} \alpha/9$,
where $A_{\rm eff}$ is the effective Hamaker constant for a metal-polystyrene interaction and $\alpha$ is the ionic polarizability~\cite{Le09}.  
In vacuum, using a standard relation
for Hamaker constants, we obtain $A_{\rm eff}^v=\sqrt{A_{pp} A_{mm}}$, where $A_{mm}$ and $A_{pp}$ are the constants
for metal-metal and polystyrene-polystyrene interaction, resulting in $A_{\rm eff}^v=17.776\times10^{-20}~\rm{J}$. On the other hand if the ion is fully hydrated
$A_{\rm eff}^w=\sqrt{A_{pp}^w A_{mm}^w}$, where $A_{mm}^w$ and  $A_{pp}^w$ are the Hamaker constants for metal-metal and polystyrene-polystyrene interaction
in water~\cite{Ru89}, yielding $A_{\rm eff}^w=6.245\times10^{-20}~\rm{J}$. 
A chaotropic ion near a latex surface, however, 
is only partially hydrated, so that the effective Hamaker constant should be intermediate between $A_{\rm eff}^v$ and $A_{\rm eff}^w$, 
which we take to be the arithmetic average, $A_{\rm eff}=\left( A_{\rm eff}^w + A_{\rm eff}^v  \right)/2$.

As the ion moves away from the surface, it creates an additional cavity from which water molecules are excluded. 
This, once again carries a hydrophobic free energy cost.  Since the volume
of this new cavity is small we can, once again, use the volumetric scaling of the cavitational free energy to estimate its cost.
This solvophobic free energy will grow linearly as 
$\beta U_{sol}(x)=-\nu a_{-}^3/2 + (3/4)\nu a_{-}^2 (L-x-a_{-})$ up to the maximum separation $ 5a_{-}/3$ from the surface, after which distance
it will be zero  --- the cavitational energy will take its bulk value.  
%

Thus for chaotropic ions, in addition to the ion-image interaction,
the solvophobic and the dispersion energies must be taken into account: $U_{-}(x)=W_{im}(x;a_{-}) + U_{sol}(x) + U_{dis}(x)$.  
Finally, unlike the kosmotropes --- which are limited by their hydration --- the chaotropes can come to the colloidal surface up to one bare 
ionic radius.
The ionic radii and polarizabilities for  $I^-$, $NO_3^-$, $Br^-$, $ClO_3^-$ and $ClO_4^-$ are the
same as used in  our previous work on surface tensions~\cite{LeDo09,DoDi10}. 
The CCC's are once
again obtained by requiring zero energy barrier for the primary minimum.  
Table~\ref{tab1} compares the theoretical predictions with the experiment. 
The good agreement in this case, however, might be fortuitous considering the crudeness of our treatment of the
dispersion interaction.  In order to test that this is not the case, we have calculated the CCC for a sodium-nitrate salt 
in suspension of negatively charged colloidal particles of $\sigma=-0.061~\rm{C/m^2}$, the CCC of which is known experimentally to be  $170~\rm{mM}$~\cite{LoJo03}. The present theory predicts $163~\rm{mM}$, suggesting that the good agreement
between the theory and experiment is not a coincidence.
\begin{table}[h]
\setlength{\belowcaptionskip}{10pt}
\caption{Comparison between the calculated CCC's and the experimental values. The colloidal surface charge 
density is $+0.04~\rm{C/m^2}$ from reference~\cite{PeOr10}.  Judging from the spread
of experimental data, the experimental error is approximately $\pm10~\rm{mM}$. }
\begin{tabular}{|c|c|c|}
\hline
      Ions            & theory (mM) & experiments (mM)       \\
      \hline
      $IO_3^{-}$      & 79.5  & 85~\cite{PeOr10}                   \\
      \hline
      $F^{-}$         & 78    & 80~\cite{PeOr10}                   \\
      \hline
      $BrO_3^{-}$     & 71.7  & *                                  \\
      \hline
      $Cl^{-}$        & 70    & 70~\cite{PeOr10}, 90~\cite{LoJo03}    \\
      \hline
      $NO_3^{-}$      & 31    & 36~\cite{PeOr10}, 35~\cite{LoJo03}    \\
      \hline
      $ClO_3^{-}$     & 25.08 & *                                    \\
      \hline
      $Br^{-}$        & 24.3  & 45~\cite{PeOr10}                     \\
      \hline
      $ClO_4^{-}$     & 20.9  & *                                   \\
      \hline
      $I^{-}$         & 14.8  & 18~\cite{PeOr10}                      \\
      \hline
      $SCN^{-}$       & 8.625 & 12~\cite{PeOr10}, 27~\cite{LoJo03}   \\
      \hline
   \end{tabular}
   \label{tab1}
\end{table}

The case of thiocyanide $SCN^-$, is somewhat more complicated since this ion can not be modeled as a sphere, but is rather a cylinder 
of radius $1.42$\AA\ and  length $4.77$\AA~\cite{IwKa82}. 
The transverse polarizability of $SCN^-$ is taken to be $3.0$\AA$^3$~\cite{PeSa05}.  It is energetically favorable for this ion to
be adsorbed flat onto colloidal surface.  The cavitational energy can be calculated similarly as above.  
With these parameters we find that $SCN^-$ is very strongly adsorbed at the colloidal surface, see Table~\ref{tab1}.  
Finally, we note that for  negatively charged hydrophobic surfaces, the CCC's follow the reversed Hofmeister series. For example,
for a suspension of particles with  $\sigma=-0.04~\rm{C/m^2}$ the CCC's range from  $70~\rm{mM}$ for $IO_3^-$ to $239~\rm{mM}$ for  $SCN^-$.  Unfortunately
no experimental data is available in this case.

We have presented a theory which allows us to quantitatively predict critical coagulation concentrations.  For positively charged
particles, the CCC's follow the Hofmeister series.  The only significant deviation is $Br^-$.  If the experimental data is correct,
this would suggest that the values of polarizability for halogen ions quoted in the literature are overestimated.  This
seems to be consistent with the recent conclusions of {\it ab initio} simulations~\cite{BaLu10}.  
For negatively charged particles the Hofmeister series is reversed. In view of the success of the present
theory it is now reasonable to hope that a fully quantitative understanding of the Hofmeister effect for protein solutions might be
in sight.  

This work was partially supported by the CNPq, INCT-FCx, and by the US-AFOSR under the grant FA9550-09-1-0283.



\bibliography{ref.bib}


\end{document}